# A Nano-Visualization Software for Education and Research

Lillian C. Oetting

Department of Computer Science, Stanford University, Stanford, CA 94305 USA

West High School, Iowa City, IA 52246 USA

Tehseen Z. Raza

Department of Physics and Astronomy, University of Iowa, Iowa City, IA 52242 USA

Hassan Raza

Department of Electrical and Computer Engineering, University of Iowa, Iowa City, IA 52242 USA

Centre for Fundamental Research, Islamabad, Pakistan

**Abstract:**

We report the development of a user-friendly nano-visualization software program which can acquaint high-school students with nanotechnology. The visual introduction to atoms and molecules, which are the building blocks of this technology, is an effective way to introduce the key concepts in this area. The software's graphical user interface enables multidimensional atomic visualization by using ball and stick schematics. Additionally, the software provides the option of wavefunction visualization for arbitrary nanomaterials and nanostructures by using extended Hückel theory. The software is instructive, application oriented and may be useful not only in high school education but also for the undergraduate research and teaching.

**I. Introduction:**

The ability to accurately depict atomic, molecular and electronic structures has been a key factor in the advancement of nanotechnology. In this context, it is imperative to provide teaching and research platforms to motivate students towards this novel technology [1-3], while keeping the societal implications in perspective [4]. Nano-visualization appeals well due to its simplistic, yet elegant approach towards the visual representation of detailed concepts about quantum mechanics, quantum chemistry and linear algebra. Additionally, the conflux of quantum mechanics, numerical computation, graphical design, and computer programming gives exposure to the multi-disciplinary aspect of this technology [5-8]. While working through various mathematical and physical constructs and visualizing them concurrently, a well-connected and application oriented picture emerges; which is in coherence with the earlier studies that visualization helps students better understand mathematics and science disciplines [9]. In our experience, a high school student may grasp these challenging concepts with ease under appropriate guidance.

In this paper, we have two objectives. The first is to report our findings about how high school students are introduced to nanotechnology by using nano-visualization software. The second is broader dissemination by offering the software for research, education and outreach.

**II. Background:**

While various visualization tools exist to aid in the process of atomic and wavefunction visualization [10-21], the reported software is novel in its integration of extended Hückel theory (EHT), for the electronic structure in wavefunction visualization, with the graphical libraries of MATLAB. EHT is a LCAO (linear combination of atomic orbitals) based semi-empirical method to calculate the electronic structure very efficiently, while capturing the essential physics. The choice of EHT and MATLAB ensures the ease of computation, programming and visualization. Therefore, the student is not distracted by tedious software debugging.

This multidimensional nanostructure visualization program is packaged with a GUI (graphical user interface), offering a level of simplicity appealing to the general user. The software has two independent modules. The atomic visualization module produces the ball and stick schematics. The wavefunction viewer module is used for plotting the atomic and molecular wavefunctions by using the EHT method. We anticipate that this software will not only be beneficial in high-school education, but also for undergraduate research and teaching.

EHT method has been successfully applied to study various multi-dimensional nanomaterials [11,22-35], including zero-dimensional structures ($C_{60}$, organic molecules, etc), one-dimensional structures (CNT, Si nanowires, graphene nanoribbons), two-dimensional structures (graphene, Si surfaces) and three-dimensional structures (Fe/MgO/Fe heterostructures). It is also noteworthy that the reported software has been used for the atomic and the wavefunction visualization in advanced research [36-40].

## II. Software Architecture

GUI programming is an important feature of the reported software. Fig. 1 shows the GUI of the software, which is designed to be user-friendly without compromising the functionality.

Atomic Structure Visualization Module

The basic capability of this software is the ball and stick schematic for atomic structure visualization. The task of atomic structure visualization begins by choosing "Select File" and then locating the desired atomic coordinate file using the standard dialogue box. The user may specify a cut-off bond length in $Å$ - if the atoms are farther apart from each other than this maximum cut-off length, a bond does not appear between them. The final step is to select "Generate Image" to display the ball and stick schematic of the atomic structure.

The user can also optimize the graphical display by using the editing tools located under "Edit Image". The atomic numbers of various kinds of atoms used in the structure can be specified. The atom color and size can be altered for specific elements which are identified by their atomic numbers. The user may specify an atomic structure comprising of up to five different types of atoms. The light check box controls the 3D effect in the ball and stick model. With light ON, the atoms appear spherical and with light OFF, they appear circular. Instead of automatically updating the image for any changes, the user selects "Done Editing" after selecting all the aesthetic quantities. The image can also be edited by a 3D rotation tool at the top figure menu or by right-clicking on the image and selecting *XY*, *YZ*, or *ZX* axes for specific view.

Finally, the image may be saved in various formats including fig; jpg; tiff; png; ps; eps; pdf; etc. by clicking "Save", which opens a Save Preview window containing the final image. Selecting "Exit" closes the program.

Wavefunction Viewer Module

The atomic or molecular wavefunctions for a nanostructure can be generated by using the *Wavefunction Viewer Module*. EHT is a valence electron theory where an atom is represented only by the valence orbitals as the basis set, each comprising of one or two Slater-type orbitals. For plotting the molecular wavefunction, a linear combination of the valence atomic orbitals with various weighting factors is used to construct the overall wavefunction. These weighting factors are provided in the coefficient matrix file (with .m extension) by clicking "Select Coefficient

Matrix File" in the Wavefunction Viewer module. In order to visualize the wavefunction, the user also has to provide a parameter file, containing the EHT parameters to be used in the calculations of the atomic orbitals, in addition to the coefficient matrix file that describes the hybridization of these orbitals. The coefficient matrix file (with .m extension) has the same order of atomic structure details as in the atomic coordinate file. Further details about the physics and chemistry of EHT are provided in Refs. [11, 22-39].

In the View sub-module, the user selects the plane *XY*, *YZ* or *ZX* at a corresponding offset of *Z*, *X* or *Y* respectively where the two-dimensional viewgraph will be plotted in a new window (with standard MATLAB editing tools) with or without bonds by clicking the Plot Wavefunction. The image parameters may be updated by selecting the Edit Settings from the menu bar, where the user may change the following parameters; Lower Bound (default value = *-3.5 Å*) and Upper Bound (default value = *3.5 Å*) define the range for which the atomic orbitals are calculated; Delta (default value = *0.05 Å*) is the grid spacing; Range (default value = *4 Å*) is the border around the structure and must be greater than the upper bound. One should note that a larger Delta value (coarse grid) may lead to erroneous location of bond display. The real part of the wavefunction is plotted by default. The imaginary part may be plotted by providing the imaginary values in the Coefficient Matrix File.

Input File Formats

The atomic coordinate file and molecular wavefunction coefficient matrix file may be provided in an ASCII file (.m file extension). The atomic coordinates are listed in the *XYZ* format, where each new line represents a unique atom and its corresponding information in the following order:

$AN_1$  $XCOR_1$  $YCOR_1$  $ZCOR_1$

$AN_2$  $XCOR_2$  $YCOR_2$  $ZCOR_2$

....

$AN_N$  $XCOR_N$  $YCOR_N$  $ZCOR_N$

where $AN_i$ is the atomic number and $XCOR_i$, $YCOR_i$ and $ZCOR_i$ are the corresponding atomic coordinates for $i=1,2,3 ... N$ atoms.

The molecular orbital coefficient matrix provides the coefficient for each atomic orbital associated with every atom in the coordinate file. In accordance with the atomic order, the format of the file containing coefficients for $s(l,m)$, $p(l,m)$ and $d(l,m)$ orbitals for $N$ atoms is as follows,

$s_1(0,0)$  $p_1(1,-1)$  $p_1(1,1)$  $p_1(1,0)$  $d_1(2,-2)$  $d_1(2,2)$  $d_1(2,-1)$  $d_1(2,1)$  $d_1(2,0)$

$s_2(0,0)$  $p_2(1,-1)$  $p_2(1,1)$  $p_2(1,0)$  $d_2(2,-2)$  $d_2(2,2)$  $d_2(2,-1)$  $d_2(2,1)$  $d_2(2,0)$

.....

$s_N(0,0)$  $p_N(1,-1)$  $p_N(1,1)$  $p_N(1,0)$  $d_N(2,-2)$  $d_N(2,2)$  $d_N(2,-1)$  $d_N(2,1)$  $d_N(2,0)$

It should be noted that for any non-existent orbital, the corresponding entry in the coefficient matrix file is *0*. The parameter file contains the EHT parameters [25,26] for up to five atoms, where each row contains the EHT parameters for one kind of atom.

**IV. Discussion of Results:**

The most important aspect of using this software is the exercise of computationally implementing the abstract mathematical and physical concepts associated with nanotechnology and visualizing them. This exercise not only solidifies the basic knowledge but also opens up new dimensions of critical thinking at an early stage of the student's academic career. This approach also serves as a driver for exciting high school students about science, technology, engineering, and mathematics (STEM) disciplines. To highlight the capabilities of this software, we consider a few examples in this section. In Fig. 2, the ball and stick models are shown for zero-dimensional (*0D*) Bucky ball molecule, one-dimensional (*1D*) carbon nanotube, two-dimensional (*2D*) graphene, and three-dimensional (*3D*) Fe/MgO/Fe heterostructure. Benzene is a standard example for which we reproduce the widely accepted wavefunction distributions for the degenerate HOMO (highest occupied molecular orbital) and LUMO (lowest unoccupied molecular orbital) energy levels, as shown in Fig. 3. For graphene, the wavefunction at the Γ point is shown in Fig. 4. The wavefunction distribution for the valence band has the same polarity as shown in Fig. 4(a). For the conduction band, however, the wavefunction has opposite polarity for the two atoms within the unit cell, which is depicted in Fig. 4(b).

We believe that such a software would find utility not only in the high school classroom environment, but also in undergraduate courses and research. The exploration of the use of the reported software in classroom trials is left as future work.

**V. Conclusions**

Combining the advanced graphic visualization libraries of MATLAB with EHT has made the reported software a unique atomic, molecular and wavefunctions visualization tool. We envision that such an effort can not only serve as an enabling platform to motivate high-school students towards nanoscale science, engineering and technology, but can also be expanded to undergraduate nano-curriculum development and research [41,42]. The software is available free of charge for research and educational purposes [43].

**Acknowledgments**

L. Oetting would like to acknowledge useful discussions with P. R. Haugen. This work was supported by the University of Iowa, Iowa City USA, and the Associateship program of the Abdus Salam International Center for Theoretical Physics, Trieste Italy.

**Figure Captions:**

Figure 1. Graphical User Interface. The user may visualize ball and stick models of multidimensional nanomaterials and nanostructures; as well as the molecular wavefunctions using the extended Hückel theory. The atomic structure visualization module works independently from the wave function viewer module.

Figure 2. Multidimensional atomic visualization. Ball and stick models are shown for (a) 0D Bucky ball, (b) 1D carbon nanotube, (c) 2D graphene, and (d) 3D Fe/MgO/Fe heterostructure.

Figure 3. Benzene molecule. (a,b) The degenerate HOMO and (c,d) the degenerate LUMO wavefunctions are shown. Delta = 0.05 Å.

Figure 4. Graphene. The wavefunction distribution at the $\Gamma$ point [$(k_x, k_y) = (0, 0)$] for the (a) valence and (b) conduction bands are shown. Delta = 0.05 Å.

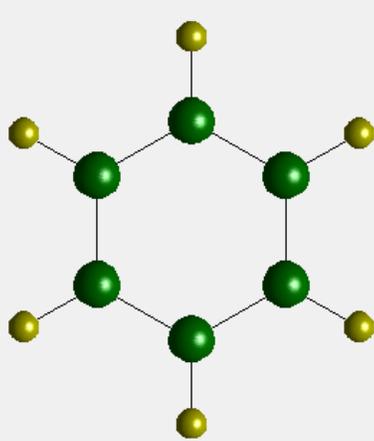

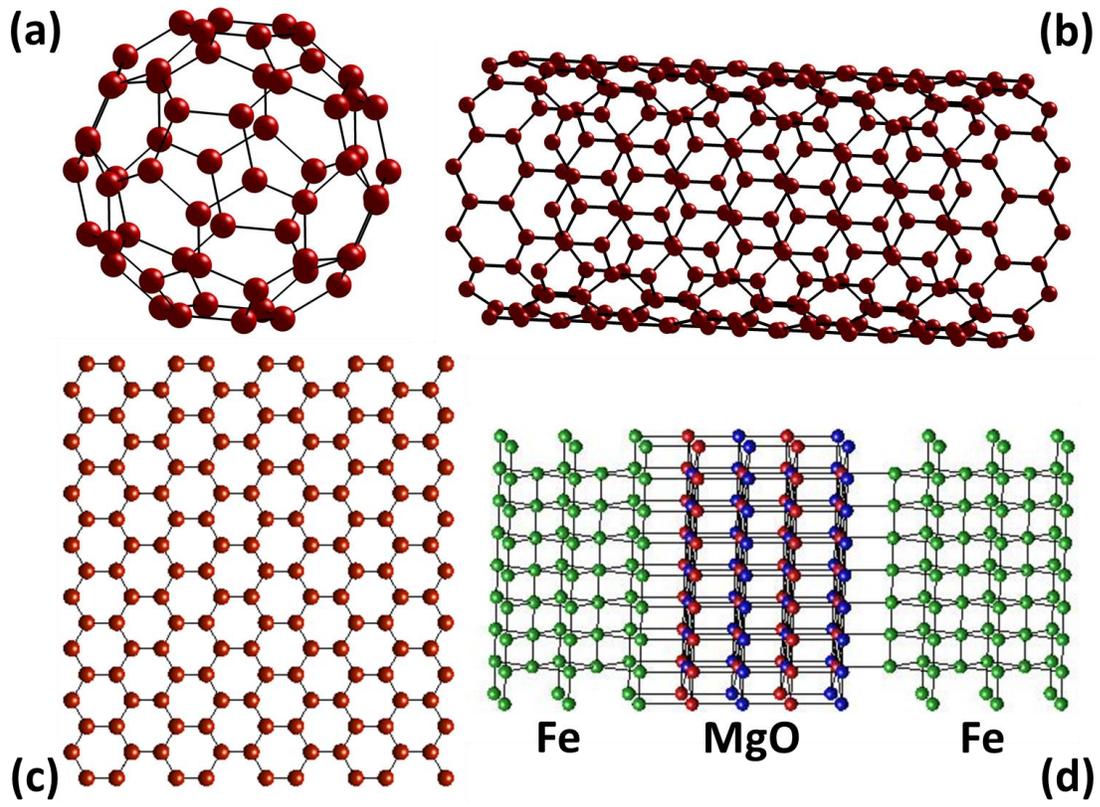

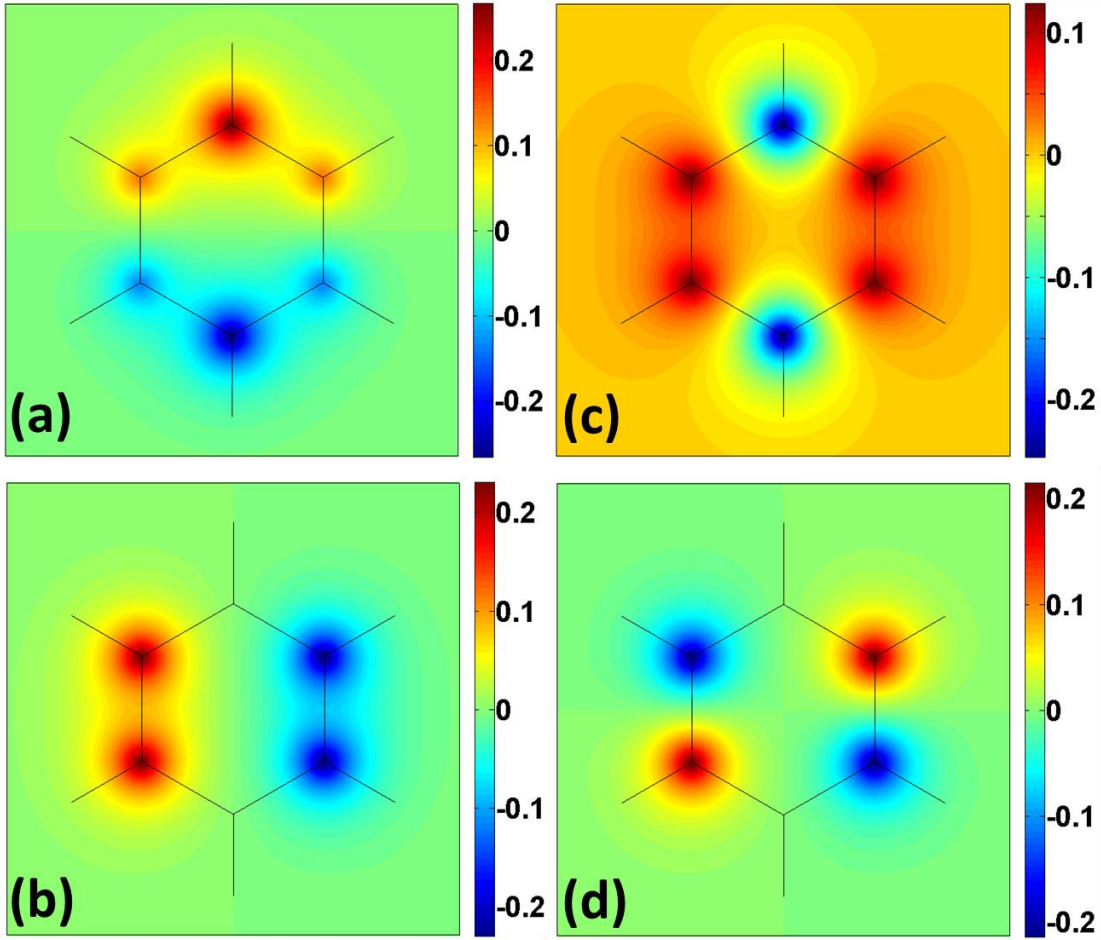

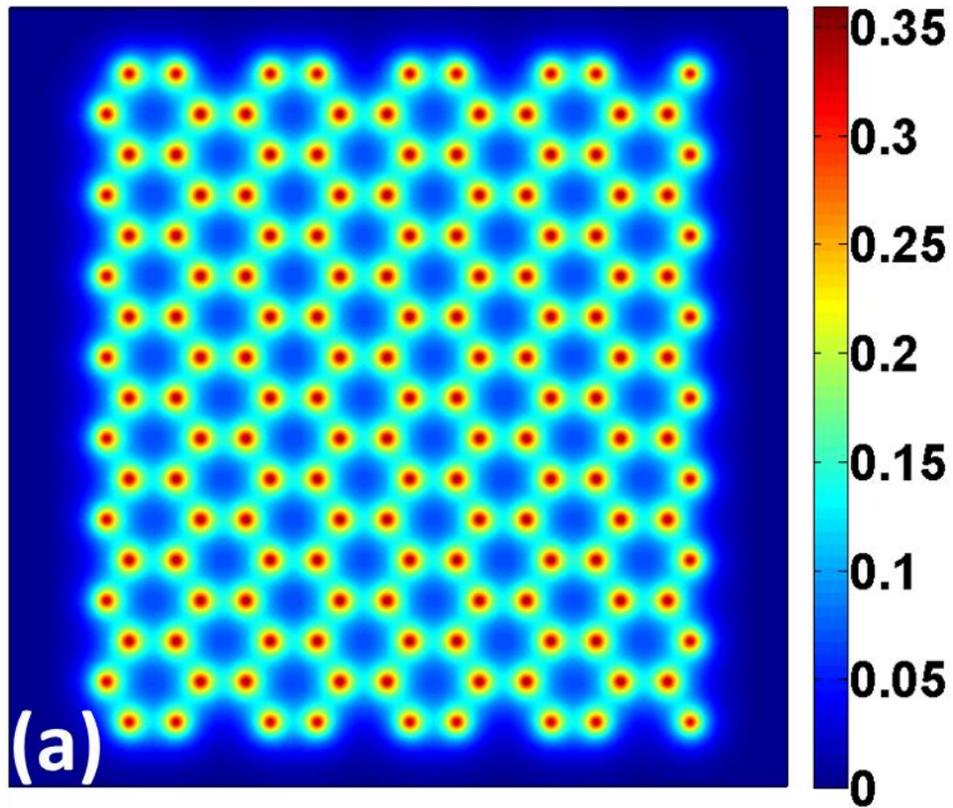

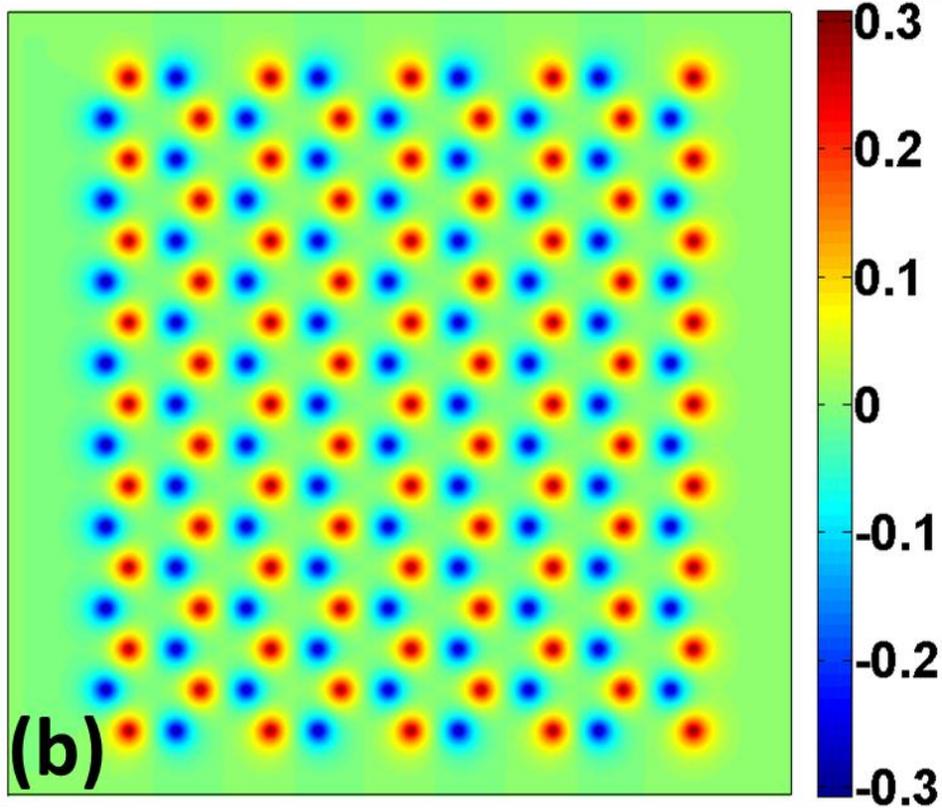